# Fabrication of sharp atomic force microscope probes using in-situ local electric field induced deposition under ambient conditions

Alexei Temiryazev[1, a)], Sergey I. Bozhko[2], A. Edward Robinson[3], and Marina Temiryazeva[1]

[1]*Kotel'nikov Institute of Radioengineering and Electronics of RAS, Fryazino Branch, 141190 Fryazino, Russia*

[2]*Institute of Solid State Physics, RAS, 142432 Chernogolovka, Russia*

[3]*AIST-NT Inc. 359 Bel Marin Keys Blvd., Suite 20 Novato, CA 94949, USA*

We demonstrate a simple method to significantly improve the sharpness of standard silicon probes for an atomic force microscope, or to repair a damaged probe. The method is based on creating and maintaining a strong, spatially localized electric field in the air gap between the probe tip and the surface of conductive sample. Under these conditions, nanostructure growth takes place on both the sample and the tip. The most likely mechanism is the decomposition of atmospheric adsorbate with subsequent deposition of carbon structures. This makes it possible to grow a spike of a few hundred nanometers in length on the tip. We further demonstrate that probes obtained by this method can be used for high-resolution scanning. It is important to note that all process operations are carried out in-situ, in air and do not require the use of closed chambers or any additional equipment beyond the atomic force microscope itself.

## I. INTRODUCTION

Currently, the vast majority of measurements made with atomic force microscopes, are carried out using silicon probes. The probes are consumable items and their lifetime is rather short, especially if high spatial resolution is required. Both mechanical wear of the tip and

---

a ) Author to whom correspondence should be addressed. Electronic mail: temiryazev@gmail.com

contamination of the tip may necessitate a probe change. In this paper we show how, using simple techniques, one can grow a thin whisker on end of the silicon tip. This allows not only restoration of the probe performance, but can actually significantly improve its performance. In particular, in this way, you can create quite a sharp probe.

Several methods have been employed to create thin spikes on top of a silicon tip for high resolution imaging. They include electron- and ion-beam induced deposition [1-3], focused ion beam etching [4, 5], CVD growth of carbon nanotubes or metal nanowires [6-9] and the attachment of carbon nanotubes [10-13]. These methods all allow the creation of very sharp probes with high aspect ratio, but the difficulties of their production and the resulting high price when produced commercially limit their application. In the case of the most widely used probes, carbon spikes are grown by a CVD process using an RF plasma in presence of Ar, $CH_4$, and $H_2O$ [14]. This process leads to the growth of several differently-oriented spikes which limits the use of probes. However, when scanning relatively smooth surfaces these probes have proved to be very effective; the high sharpness of the spike, with a radius of curvature of about 2 nm, allowing high resolution to be achieved in ambient conditions. Such probes have been applied in several studies carried out by different groups [15-21].

In this paper, we describe a very simple and effective method to initiate spike growth on a silicon tip. The method does not require a closed-chamber environment nor explosive hydrocarbon precursors. We will demonstrate that such growth can occur in air, when a high electric field is created in the gap between the tip and the surface of a conductive specimen. The most likely source of the material is atmospheric adsorbate located on the surfaces of the sample and the needle. Decomposition of hydrocarbons comprising the adsorbate with subsequent deposition of the released material leads to growth of carbon nanostructures on the tip and on the sample surface. The process is similar to that occurring at the sample surface in electron microscopes if the vacuum is poor, or the chamber contaminated. By analogy with electron beam induced deposition (EBID), we can say that we are dealing with local electric field induced deposition (LEFID). Unlike EBID, the growth occurs not only on the sample surface, but also on the tip, around which the electric field is localized. This process can be used to create the sharp probes. The spatial resolution that can be obtained with such probes is comparable to the best measured to-date under ambient conditions. This paper consists of two parts. In the first part, we consider experiments demonstrating the possibility of LEFID under ambient conditions. The second part describes the process of tip manufacture and give examples of their use to obtain AFM images with high resolution.

## II. EXPERIMENT AND DISCUSSION

All studies were performed using a SmartSPM atomic force microscope (AIST-NT). For surface scanning we used a dynamic mode, i.e, the probe is driven to oscillate at a frequency close to resonance, and the amplitude of oscillation is used to maintain a constant tip-sample separation as the sample is scanned under the tip, with the z-piezo position thus mirroring the surface topography. Scanning was carried out either in semicontact (tapping) mode, or in the case of rough samples, vertical mode (VM). In the latter case, the height measurement in each point is performed during probe approach at a constant speed, with the sample X/Y position held constant. The measurement of the height is made when the amplitude falls to a set level, $Am_1$. The probe is then raised and the horizontal displacement maintained until the amplitude of oscillation exceeds $Am_2$, where $Am_2$ is 5-10% above $Am_1$. The advantages of VM result from the low forces between the probe and sample (because it is a dynamic mode, and can operate at a low amplitude of oscillations), the absence of feedback generation near vertical walls (since PID control is not used), and relatively fast scanning of smooth regions (as $Am_2$ only slightly exceeds $Am_1$). More details of VM can be found in [22]. For measurements requiring high spatial resolution we used dissipation mode [23]. Below, we take a closer look at this latter mode.

In order to initiate growth of the desired spike on the tip, we applied voltage $U$ to the sample through a 20 GΩ ballast resistor. The current passing through the tip was measured with a local preamplifier (included in the microscope), connected close to the probe. The tip was brought into tapping mode feedback (half the peak-to-peak value of the free amplitude about 15 nm , set point 70 %); a voltage was applied to the tip and slowly (over a period of about 1 s) increased to between 30 and 100 V. As the voltage rises, the resonant frequency of the cantilever shifts, the amplitude drops and the tip moves away from the surface. AFM feedback was then turned off and current feedback was established (similar to scanning tunneling microscopy (STM)). A typical set point current is about 1 pA. The conventional AFM laser beam deflection signal is simultaneously monitored (similar to contact mode AFM) in order to prevent the tip from crashing and damaging the sample and/or the probe, in case of a contaminated or non-conductive area of the sample. If mechanical contact between the tip and the surface causes a change in deflection of the cantilever, further movement of the probe into the sample is arrested, even at zero tip current. We used two modes of operation: one a discrete point mode where, after a few seconds, the probe is retracted from the surface and moved to another point, the current feedback turned back on and the process is repeated; the other a line mode, where the sample is moved at a speed of 10-300 nm/sec while maintaining current feedback. In each case, the probe is then retracted, moved to the start of the next line and the process resumed. At the end of the operation, the voltage was removed and the surface scanned in dynamic mode. All experiments were performed in air. The samples were semiconductor (Si or GaAs) or metallic (nickel, cobalt, gold, platinum) thin films., with most of the work performed on nickel films. No special treatment of the surface was undertaken. We used probes from different vendors (mostly NSC15, MikroMasch and fpn10 or fpn1, AIST-NT) with a nominal stiffness of 3 to 40 N/m. Some probes had metallic conductive coating, but the presence of such a coating is not essential.

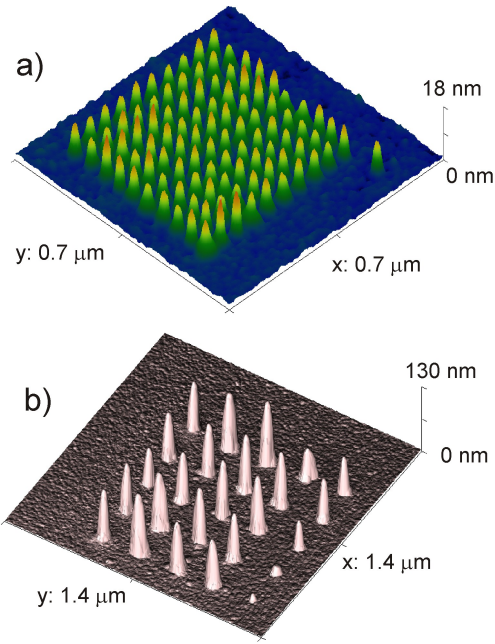

FIG. 1. Point deposition on the surface of Ni film. (a) Bias voltage $U = 30$ V, time per point $t = 2$ s, set point current $I = 1$ pA. (b) $U = 54$ V, $t = 20$ s, $I = 0.5$ pA.

Now we consider the results of some experiments, which demonstrate the modification of both the sample surface and of the tip resulting from an applied voltage. In the first, a spike is created on the sample surface if we apply voltage at a point. Moving from point to point, we can create an array of spikes. Figure 1 shows two such arrays. In the first case, the period is 50 nm, and a height of about 10 nm (Figure 1(a)). It can be seen that the resolution of the method is rather high. In the second case (Figure 1(b)) period is bigger, 200 nm, and the spikes are substantially taller. The tallest has a height of 110 nm with a measured width (FWHH) of 60 nm. Given that the measured feature is the convolution of the object shape with that of the probe tip, we know that the sum of the widths of the tip and the substrate spike is 60nm and that the aspect ratios are quite high.

In the second method, moving the sample under the probe, a raised line appears on the surface under the influence of the electric field. Thus, a raised pattern is created on the surface (Figure 2). The resolution is quite high, the width of the lines shown in Figure 2(a) is 20 nm with a height of 12-15 nm. It should be noted that parameters of line are not always stable, both smooth and abrupt change in width and height are seen. The lines near the eyes in Figure 2(b) are noticeably narrower, resulting from line instability rather than intentional change of parameters.

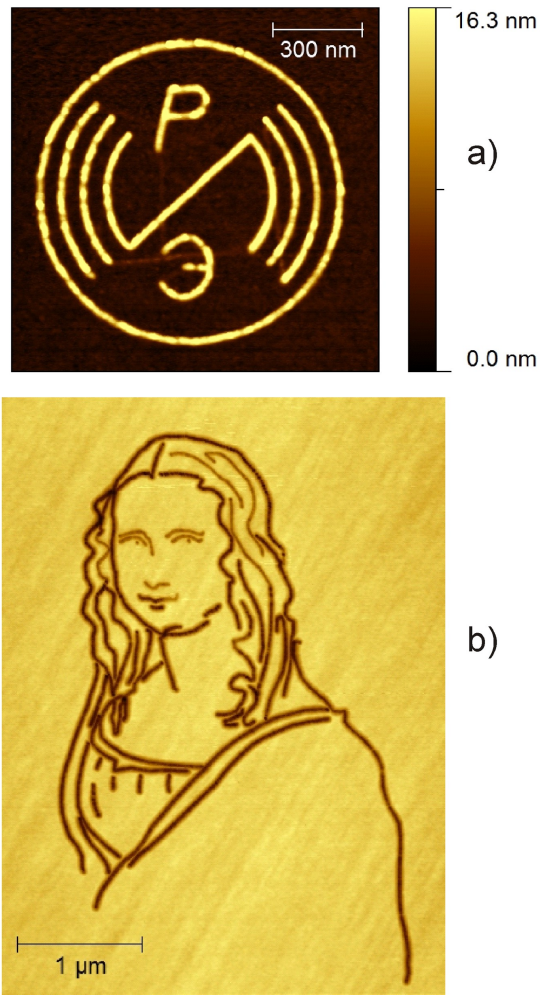

FIG. 2. (a) Line pattern deposition on the surface of Ni film. The line height is 12-15 nm, width – 20 nm. (b) Line pattern on the surface of GaAs, colors are inverted, the line height is about 8 nm. The drawing is based on a colouring page from http://www.kleurjeslim.be.

In general, it should be noted that height and width depend on many, generally speaking, unknown parameters. Some of them are simple enough to control - voltage, current, speed of moving. Some - temperature, humidity, pressure, gas composition environment, contamination of the surface could be changed using a closed chamber. We conducted experiments in the open air, so such studies are yet to be performed. The effect probe shape is
6

especially difficult to account for because the tip is continually growing as further material is deposited on the substrate. It is likely that the resulting change in the shape of the probe is a significant source of line instability.

The shape of the probes used for line deposition was studied with scanning electron microscopy (SEM). Spikes several hundred nanometers in length were found at the tip end (Figure 3). The isotropic shape in X/Y of the spikes imaged on the substrate in Figure 1(b) leads to the conclusion that both substrate spike and probe spike are formed perpendicular to the plane of the substrate. So, it is not immediately obvious why the spikes imaged in the SEM are not close to the expected 20 degree inclination due to the cantilever geometry in the SmartSPM head. The observed thickness of the spikes is also quite large, a few tens of nanometers. This value is probably an overestimate, due to overgrowth of the probe surface during the SEM measurements. An assumption supported by much smaller widths of the broken filaments sometimes seen in subsequent AFM images of the surface. Consider in more detail one of these experiments.

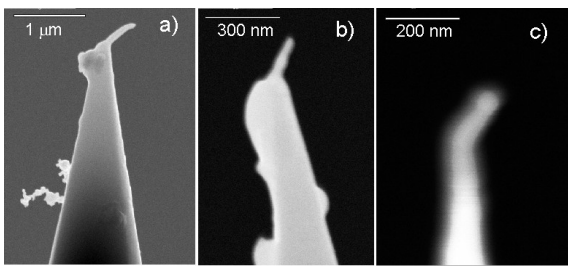

FIG. 3. SEM images of three different tips with the grown spikes.

A region of bare substrate was scanned and determined to be flat and featureless. Varying the voltage and moving speed, we then deposited a series of three parallel, one-micron lines, oriented at 45 degrees. Three such series can be seen in the lower part of Figure 4(a), reveled in subsequent scanning; additional objects are also detected. A more detailed study of one of these objects (Figure 4(b)) reveals a fairly straight structure with a height of about 10 nm and a length of several hundred nanometers. We assume that this is nanowire or nanotube lying on the surface. Most likely, it grew up at the end of the tip, then broke off; the doubled image is obviously due to a defect in the probe which, at the time of this scan, apparently bore two spikes. We then tried to improve the imaging, hoping to break off one of the spikes by pressing the probe to the sample surface. The scan made after this (Figure 4(c)) shows that the probe has actually become very blunt. Such a probe has little value for scanning, but it can be used to see how the nanowire responds to mechanical impact. So, we again pressed the probe to the sample and began to move across the surface in contact mode, to shift the nanowire sideways. A subsequent scan (Figure 4(d)) shows that the structure has both shifted and become bent. The next task was to restore our probe. We moved this blunt probe away from the scanning area and applied voltage for 20 s at a point, building up a fresh spike on the tip. A re-scan of the object immediately afterward shows (Figure 4(e)) that the operation was successful, a clear image again being obtained. The height of the nanofilament is 9.5 nm with the apparent width of 22 nm. Its length is greater than 800 nm.

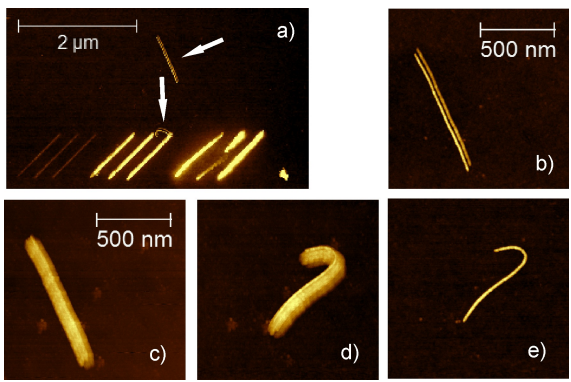

FIG. 4. (a) AFM image of the surface of Ni film after the deposition of 9 lines. Two additional unexpected objects (straight and curved) marked by white arrows were found. (b-e) Subsequent AFM images of the initially straight object.

We have used imaging performance, without resorting to SEM imaging, to demonstrate that spikes with length of several hundred nanometers can be grown on the tip. To do this, we started with a badly damaged silicon probe. To determine the shape of the probe, we scanned test grid TGT 1 (NT-MDT), which bears a regular, engineered array of sharp features, 300-500 nm in height. Scanning this test sample with a super-sharp probe having a high aspect ratio will provide an image close to the shape of the grating feature. For a case of a very blunt probe tip, scanning will show the form of the probe tip, effectively imaging the tip with the sample. In the intermediate case, the shape of the probe may be calculated by mathematical deconvolution of the image, indeed this is the purpose of such a test sample. Figure 5 shows a quasi three-dimensional images of TGT 1 obtained by scanning with a damaged probe and then with the same probe, after growth of a spike to restore its utility. It is evident that we have succeeded in restoring the damaged probe.

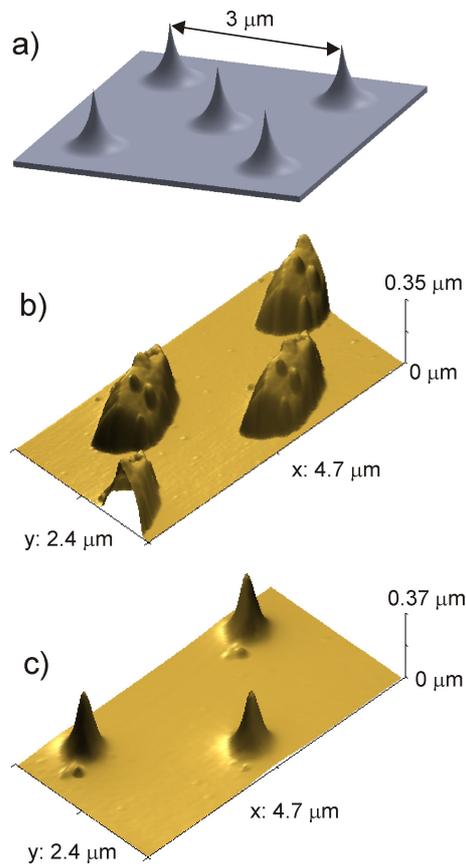

FIG. 5. (a) Design of test structure TGT 1. (b, c) AFM images of TGT 1 obtained using a heavily-damaged probe tip (b) and using the same probe after subsequent spike growth (c).

After deposition, the probe has a sharp tip of a length of about 350 nm. To show that this change is not simply related to removing particles adhered to the probe, Figure S1 in the supplementary material shows also some intermediate results of the measurements. Initially, the probe had a protrusion of 50 nm height. After the first growth procedure, scanning the testing sample showed that the height of the protrusion had increased to 100 nm, with the blunt shape of the original probe still seem beyond this level. It was only the next operation (the deposition of a series of lines on the surface of the sample) that resulted in a spike long enough to image the entire height of the spike array without detecting the supporting blunt tip.

It should be noted that modification of the surface by an electric voltage applied to the AFM or STM tip may be caused by several different mechanisms [24]. For example, similar effects occur by local anodic oxidation (LAO) [25, 26]. The applied voltage for LAO is somewhat lower, typically less than 20 V and the current, in the cases when it was controlled, is in the range of pA to nA [27-29]. We do not exclude that anodic oxidation can occur in our experiments, but this process is not determinative. LAO is a self-limiting growth process [30]. The height of the oxide formed during LAO is small, a few nm, and the width is substantially greater than the height. In our case, where the deposition is on platinum or gold surfaces, tall spikes with a large aspect ratio are formed even on noble metal surfaces, where no oxidation would be expected.

Another method of surface modification is the direct transfer of material from the probe to the surface [31-37]. The distinctive feature of these experiments is that the voltage is applied in short pulses, the distance between the probe and the surface is very small, typically less than 1-2 nm. Field evaporation is recognized as the most likely mechanism, although mechanical contact of the probe with the surface was also considered [33]. In our case we are dealing with much longer time, and the small currents mean a much larger distance from the probe to the surface. Growth occurs at both the probe and the sample, suggesting that there is an additional source from which the material is deposited.

The closest analogue of our experiments seems to be local CVD growth by STM [38-42]. A variety of gaseous, organometallic precursors were used in such experiments and deposition both on the sample and simultaneous growth of a filament on the tip were observed [40]. In our case, the experiments were conducted in air, without exogenous precursors. The most probable mechanism of deposition is decomposition of hydrocarbon compounds present in the air and adsorbed on the sample surface. The growth of surface contamination in the beam is very common for electron microscopy, generally the scanned area can be clearly seen in subsequent lower-magnification scans as a raised rectangle. As for AFM, the presence of a carbon residue in oxide patterns formed under ambient conditions was reported in [43].

In a recent paper [44], field-induced-growth of a nano-whiskers on a silicon tip apex was attributed to crystalline-to amorphous phase transformation on the surface of the silicon nano-cathode. The presence of a diamond-like-carbon coating suppressed this process. In our case, the growth was observed both when using silicon tips and with coated (Pt, $W_2C$) probes, therefore such a mechanism is unlikely.

So, we believe that we have devised conditions under which one can maintain a large electric field in a confined area, near a surface. Spatially, this field is extremely heterogeneous, since the second electrode is a tip of the probe. The drift of surface adsorbate in an inhomogeneous field, its decomposition and subsequent self-organization of fragments appear to be the most probable mechanism of spike formation.

For those interested in applying these techniques of AFM tip improvement, we make some key observations:

Firstly, the large voltage difference between the probe and the sample can lead to electrical breakdown. If uncontrolled, the result will be significant damage to both probe and sample [45].

In our experiments, the large ballast resistor limits the power of discharge, therefore only in very rare cases was the formation of pits on the surface of the sample observed, and only at higher voltages (about 100 V).

Second, maintaining a constant current level is not always possible, perhaps due to the presence of an additional degree of freedom, such as bending of the cantilever due to attraction to the surface by electrostatic forces. Occasionally, as the tip approaches the surface, sudden current increase occurs, current-controlled feedback withdraws the probe, the current declines, approach starts again and the process repeats. Usually, after a series of fluctuations, the current stabilizes.

Third, air humidity is an important factor: at an RH more than 60%, protuberances on the surface become lower and broader and no noticeable growth of a whisker on the tip is observed.

Fourth, most of our experiments have been conducted with a positively-biased sample. When the polarity is reversed, the formation of bumps on the surface is much less pronounced. This regime has not been studied in detail.

Fifth, the process of growth proceeds by self-organization, and there are several possible scenarios of its development. To illustrate this point, consider the results of another experiment, which also serves to give an idea of typical growth rates. We recorded the increase of probe height with time, while maintaining current-controlled feedback. Figure 6(a) shows the plots. The sample was biased at +48 V. The dwell time of the probe was 10 seconds at each of the 3 points. At the first point (time segment AB), we see a fairly slow increase of height. Apparently, material is being deposited both on the sample and on the probe. At time B the probe retracts and moves to the next point of the sample. At the second point (time segment CD), the height increases very rapidly, but at the time D it drops sharply, apparently the grown spike breaks. At time E the probe retracts again and moves to the third point. Here, after a relatively slow change in the height during interval FG, there is new fall, the height does not change for a while and then it begins to increase rapidly in the area HI. The plot in this case recalls the initial portion of the segment CD.

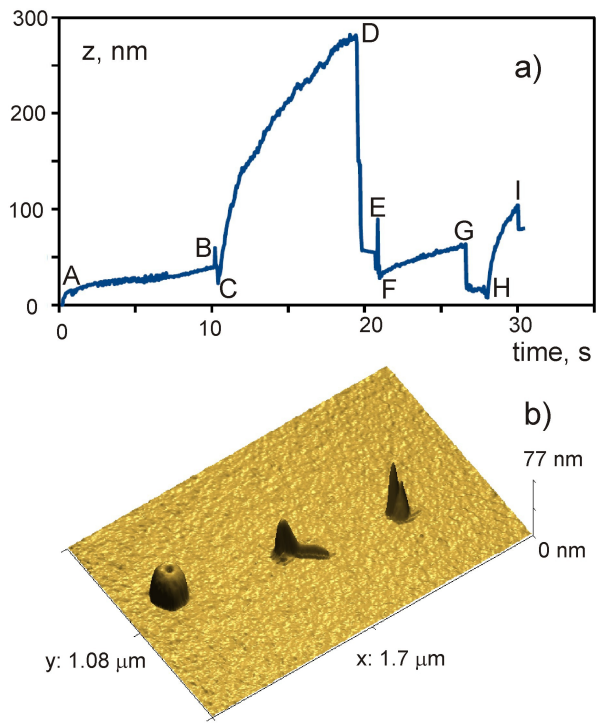

FIG. 6. (a) The dependence of the height position of the probe over time, when a tip biased at -48 V was located (I = 1 pA) for 10 s in three different points on the surface of Ni film. (b) AFM image of the surface after this operation.

The scan (Figure 6(b)) shows the structures that were formed as a result of this operation. At the first point (time segment AB), we can see a mountain with a crater in the center (the height of 38 nm, 140 nm diameter). At the second point, there are a narrow spike of 49 nm height and, lying on the surface, a filament of about 180 nm length, height of 15 nm, and apparent width of 25 nm. At the third point, there is a narrow 75 nm peak, on the background of a broader mound with height of about 40 nm (clearly seen in Figure S2(a) in the supplementary material). Interestingly, the phase image (Figure S2(b) in the supplementary material) shows distinctive phase contrast in quite a wide area around the points of application of the electric field. This difference in surface properties is perhaps caused by the drift of adsorbate into the area of high field.

### III. MANUFACTURING SHARP PROBES IN-SITU

Thus far, we have shown that by applying a voltage, deposition occurs both on the sample and at the probe tip. Most likely, the material is carbon, resulting from the decomposition of hydrocarbon adsorbate. The next task - show that this effect can be used for practical purposes, namely for the manufacture of a usable, sharp AFM probe. There may be two extremes to the requirements. The first is to provide probes that allow utmost lateral resolution in air, the other to allow accurate visualization of features with significant topography.

In the first case, a lateral resolution on the order of 1 nm requires probes with sharp tip. The shape of the probe above the tip is not so important because most of these probes are used for fairly flat objects; in this case, the tip radius characterizes the probe. Standard silicon cantilevers have a radius of 10 nm, commercially-available super-sharp tips with a carbon spike have a radius of about 2 nm [14].

In the second case, the visualization of nanoparticles or topographies with relatively large height differences (30-100 nm) and narrow troughs, tip radius is not as important as the aspect ratio of the probe above the tip.

Using the technology introduced here, we can solve either of these problems.

Though the growth mechanism may not be completely understood, it is apparent that we may quickly evaluate the performance of any given tip and, if it is not suitable, immediately repeat the operation to achieve the desired result, by adjusting the parameters.

Let us see how this is implemented in practice: To obtain high-resolution, we need a tip with a small radius of curvature. One can estimate the radius by using the same AFM, e.g., with the method suggested by Santos et al. [46]. This method is based on the fact that the interaction of the probe with the surface depends on the sharpness of the tip. The blunter the tip, the larger the area of interaction and the bigger the contribution of the attractive forces. The sharper the tip, the lower the amplitude of free oscillations $A_0$, that provides a regime in which repulsive forces prevail over the forces of attraction. As will be seen, the criterion for such a regime is a negative phase shift (as a result of an increase in the resonant frequency of the probe interacting with the surface).

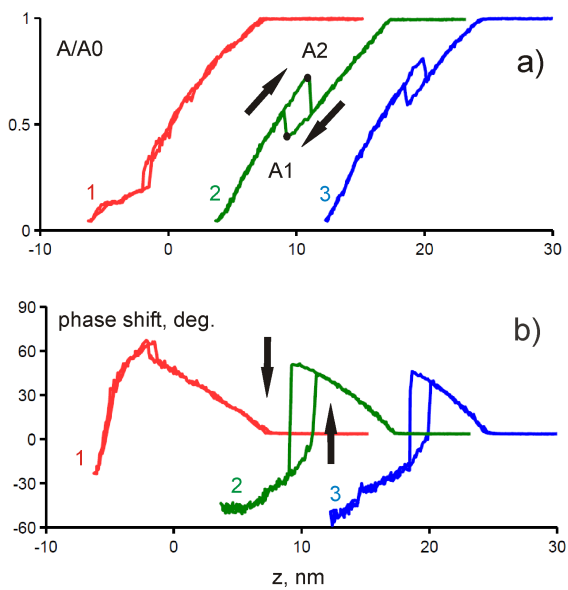

FIG. 7. Experimental approach curves for the same probe increasing its sharpness from curve 1 to curve 3. Approach was conducted in tapping mode; that is, at constant oscillation frequency. For clarity, curves 2 and 3 are offset in the horizontal axis: the $Z$ displacement.

At very low amplitude of free oscillation $A_0$ this shift is always positive. Starting from a certain critical amplitude $A_c$ of free oscillation, that is, when $A_0 > = A_c$, an approach curve has a region with a negative phase shift. The technique proposed in [46] assumes the measurement of $A_c$ during approach of the probe to a specific test surface (e.g., mica, graphite, silicon), for which numerical relations between $A_c$ and radius of curvature were obtained. This is not ideal for us, because it may require a change of the sample. But, since the precise determination of the curvature radius is not our aim here, it is sufficient to develop criteria to determine if our probe is sharp enough to be used for our purpose of high resolution scanning. Therefore, we have not attempted to determine $A_c$ but used the fact that when the amplitude is above the critical level, a sharper tip more quickly enters the repulsive regime as it approaches the surface; a very sharp probe transitions to the repulsive regime almost immediately after touching the surface.

Figure 7 illustrates this approach. There are several specific areas on the approach curves: When approaching the surface, the phase shift becomes negative at A1. During retraction, there is a reverse phase jump at A2. Thus when the amplitude $A < A_1$ we have a repulsive regime and the phase shift is negative. In the range $A_1 < A < A_2$, bistability is possible and the phase shift can take both positive and negative values. If $A_0 > A > A_2$ phase shift is positive. The sharper the tip, the higher the level of $A_1$. If $A_1$ is close to $A_0$ then $A_0$ is substantially higher than $A_c$. Thus, if the free amplitude is small and we scan with a relatively high set point while measuring a negative phase shift; it is safe to say that the critical amplitude is significantly lower than $A_0$ and that the tip has a sufficiently small radius of curvature. We used the following parameters: an initial amplitude $A_0$ of about 15 nm, set point $A_{sp}$ at $0.7A_0$. After growth, we scanned the surface and, if the phase shift on a flat area was negative ($A_1 > A_{sp}$), considered that we had generated a sharp tip, if not, we immediately repeated the operation to grow a better tip.

The three curves shown in Figure 7 result from the same probe and the same initial amplitude of about 15 nm. They differ only in the sharpness of the probe, which improves with the number of the curve, the growth procedure having been repeated in between the three acquisitions. As the tip becomes sharper, the region where phase shift is negative increases. If $A_{sp}$ was 50% of the $A_0$, during the scanning we would see positive phase shift in the case of curve 1, phase jumps for the curve 2, and a negative phase shift for the curve 3. We used a more strict condition, setting the Asp at 70% of the initial amplitude of 15 nm. The growth procedure was repeated until the scanning showed negative phase shift at this set point.

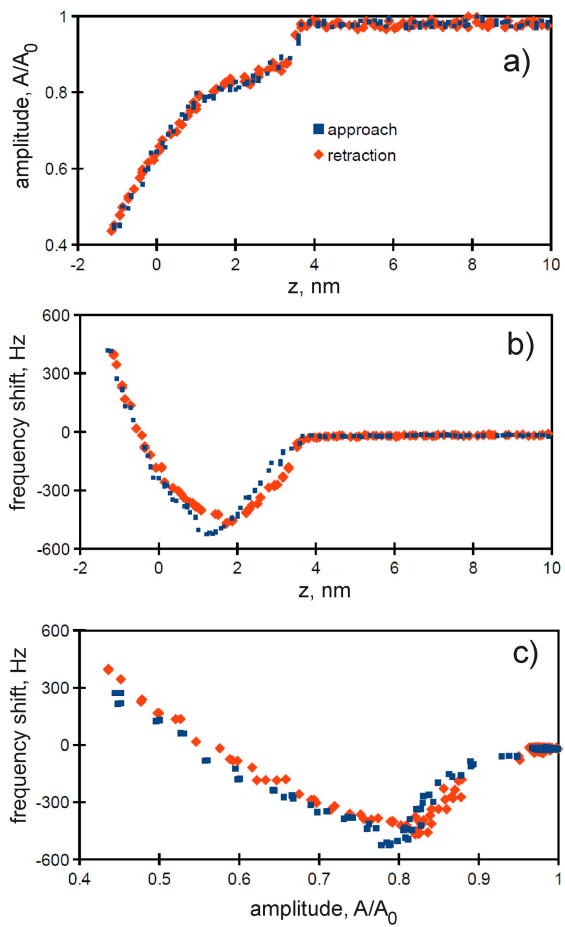

FIG. 8. Experimental approach curves in case of dissipation mode: i.e., at resonant frequency. Amplitude versus distance (a), resonant frequency shift versus distance (b), and frequency shift versus amplitude (c).

Probes produced in this way were used for research requiring high lateral resolution with measurements carried out in dissipation mode (details in [23], here we present only an overview and results from this mode). Pump power is fixed, and the oscillation amplitude serves to maintain the feedback control (as in tapping mode) and the oscillation frequency is maintained at resonance (as in frequency modulation mode). Under these conditions, a sharp probe gives us the possibility of observing reversal of the frequency curve at low free amplitude $A_0$ ( Figure 8). The amplitude at which the reversal of the frequency curve takes place is selected as the set point. From the approach curves taken in dissipation mode and shown in Figure 8, it is clear that the initial amplitude is 8-9 nm, a distinct reversal of the frequency dependence begins about 2 nm after touching and that a set point of 80% of the free amplitude is a good initial choice. Of course, $A_{sp}$ can be precisely adjusted during the scanning to optimize imaging.

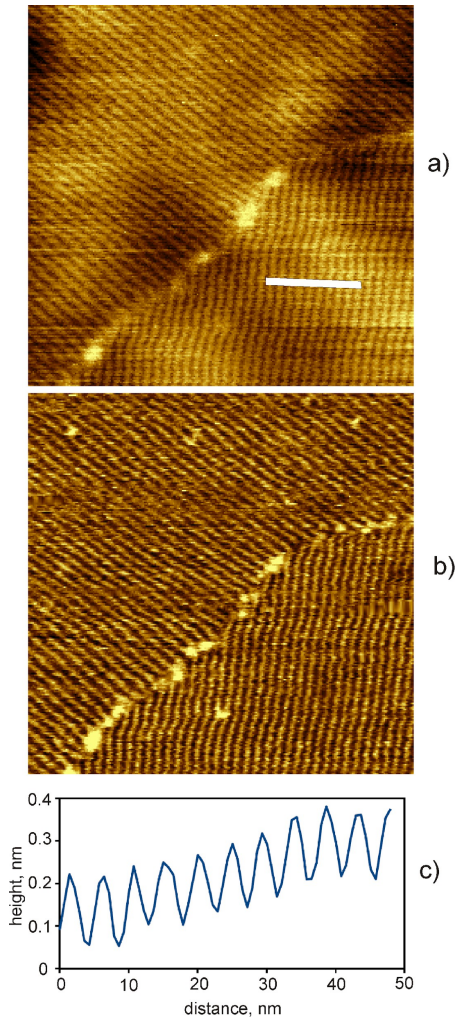

FIG. 9. Example of images obtained by using the probe with grown spike. Self-organization of adsorbate on the graphene. Scan area 180×180 nm$^2$, (a) – height, (b) - frequency shift, (c) the height profile along the white line shown in (a). Color scale in (a) is 1.4 nm, in (b) color shows the frequency variation from -340 Hz to -400 Hz.

Both resonant frequency variation and the height image were recorded, giving excellent resolution of fine details: Figure 9 shows an example of the scan obtained in this manner. Here we investigated the structure of an absorbed monolayer on the surface of graphene. The period of the observed striped structure is 5 nm. Fine details are easily distinguished in both height and frequency images, proving the sharpness of the tip. We have made scores of such probes and used them to produce thousands of scans. The tips are fairly durable: in one experiment a single probe was used to perform 100 scans at 100 different points on the sample, with no noticeable degradation of the probe. The details of this experiment will be presented elsewhere.

Our second objective, to obtain probes with high aspect ratio, can be achieved using a similar method, simply by changing the parameters. In fact, one can easily create sharp spikes of about 50 nm height on the surface, subsequent scanning of which shows the suitability of our probe for samples having such height difference. One can make arrays, such as those shown in Figure 1, and use them to estimate the form of the grown spike. The only thing to consider in this case is that the arrays in Figure 1 were made by the sharp probe, which had already bore a suitable spike. This allowed the creation of sufficiently dense arrays, with a small period. If starting with a badly damaged probe with a blunt end, one should first carry out growth on a flat area of the sample and set a larger distance between the points, on the order of 500 - 1000 nm. Otherwise, multiple protrusions may appear on the tip, though not on the sample. An example of this is shown in Figure S3 in the supplementary material.

To test the efficiency of the obtained probe, we scanned the surface of electrochemically structured (dimpled) aluminum. Such samples are often used to demonstrate probe quality, since they have thin walls and sharp protrusions at the vertices of hexagons [13, 47]. A spike was grown on a tip coated with platinum (though, as already noted, the presence of a metal is not necessary, an uncoated tip can be used just as well). The metal coating increases the tip radius; the scan of the sample with an unaltered probe shows the result expected from a common, blunt probe (Figure 10(a)). After spike growth, the tip became much sharper;

defects in the walls are now distinguishable as are the protrusions at the intersection of walls (Figure 10(b)).

FIG. 10. AFM images of the dimpled aluminum sample obtained by using the probe with the Pt

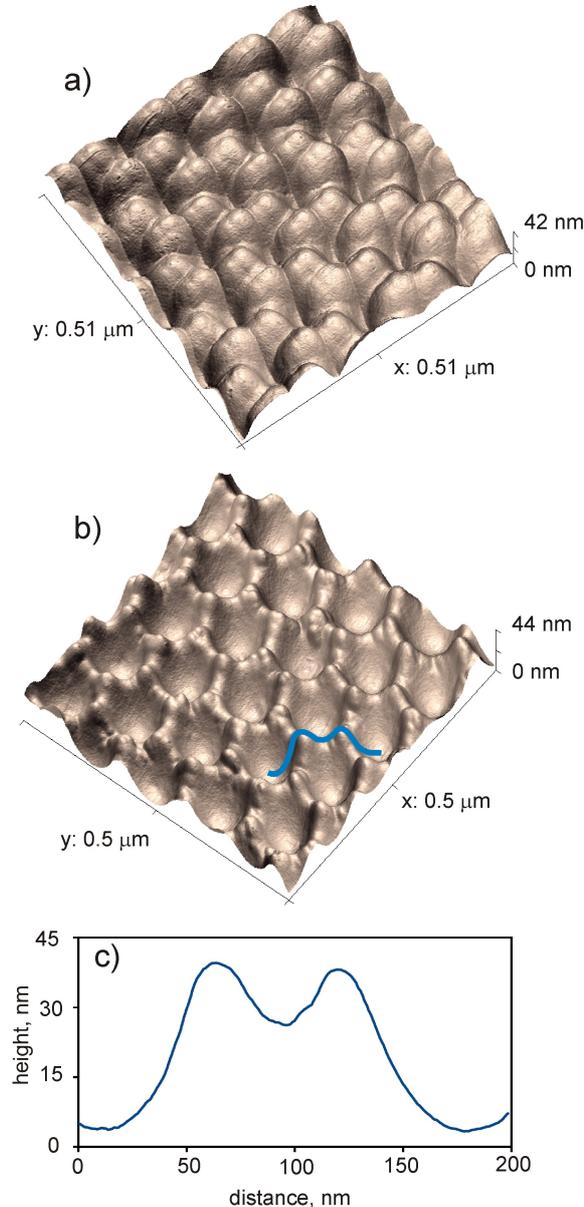

covering (a) and the same probe after whisker growth (b); (c) the height profile along the blue line shown in (b).

Next, we conducted an experiment to evaluate the durability of the probe during extended scanning. A series of scans was carried out over the same 500 nm × 500 nm area. Thermal stabilization was not applied for drift compensation; instead we used one of the protrusions on the surface as a reference point, the software holding an automatic correction of the scanning area between scans. A similar procedure has been used in the past in STM experiments [48, 49]. No change was observed over the course of the first thirty-two scans, then, during the course of the thirty-third scan, the imaging changed dramatically, apparently due to the spike having broken. A new spike was built on the same probe and the experiment started afresh. This series completed without incident; 100 scans were obtained, of which we made a movie (see Figure S4 in the supplementary material).

## IV. CONCLUSIONS

In conclusion, it should be noted that, although the specific values of the growth parameters (voltage, current, time or speed) shown in the figure captions are typical, optimal conditions vary with the environment; all processes are carried out under ambient conditions. Deeper study of the mechanisms of spike formation is necessary in order to make the process more stable and predictable. Nevertheless, our studies may be of practical interest because they significantly narrow the range of search. It has been found best to use relatively high voltages (over 25 V) and small currents (not exceeding 1-2 pA) and that it is essential to provide protection against electrical breakdown. Under these conditions, nanostructures grow both on the tip and on the surface of the sample. This allows to make sharp probes and the structures deposited on the surface can themselves serve as a mask for subsequent processing steps.

### SUPPLEMENTARY MATERIAL

See supplementary material for additional figures and a movie of the durability test.

### ACKNOWLEDGMENTS

This work was partially supported by the Program of Presidium of Russian Academy of Sciences.

Supplementary material for

# Fabrication of sharp atomic force microscope probes using in-situ local electric field induced deposition under ambient conditions

Alexei Temiryazev, Sergey I. Bozhko, A. Edward Robinson, and Marina Temiryazeva

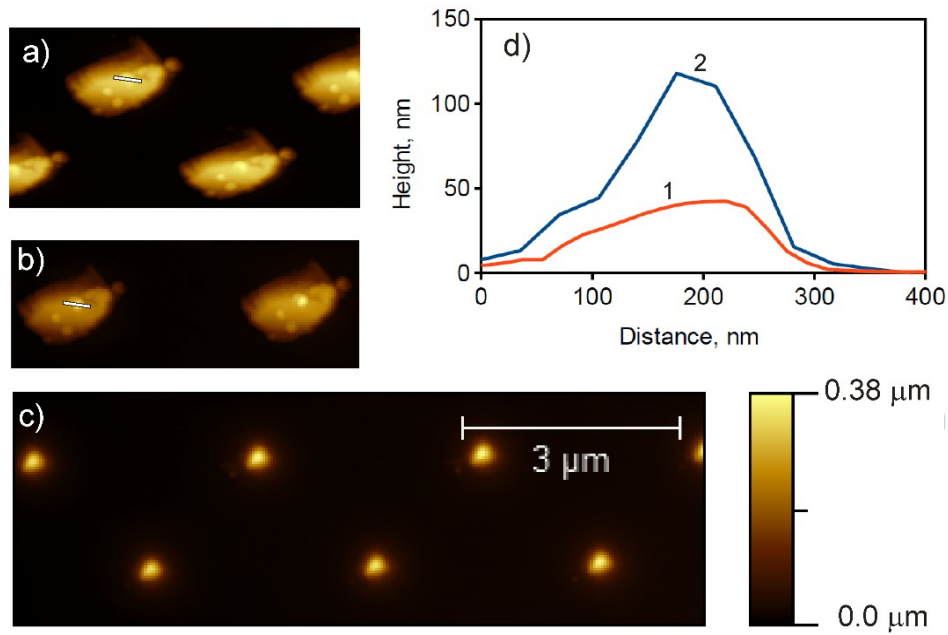

FIG. S1. AFM images of test structure TGT 1 obtained using a heavy damaged tip (a), the same probe after the first (b) and second (c) operations of spike growth. (d) cross-sections: red curve 1 corresponds to line in (a), blue curve 2 – to line in (b).

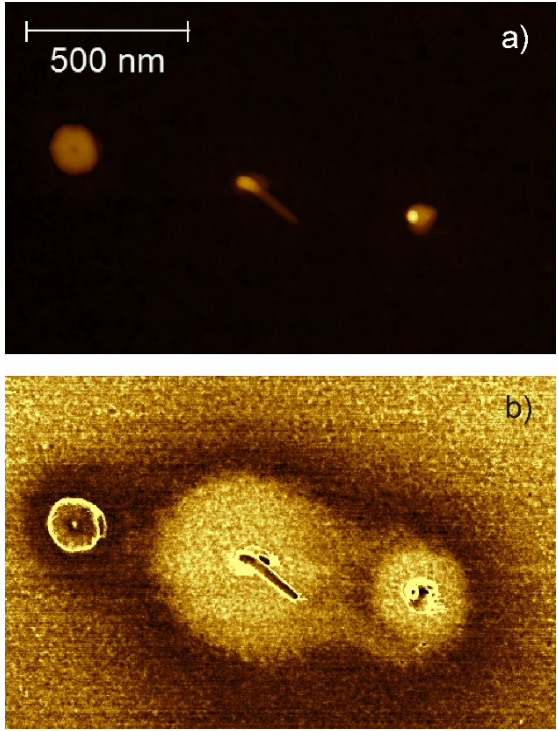

FIG. S2. AFM image of the surface of a Ni film after voltage was applied at three points; (a) height, (b) phase. The phase image clearly shows that surface properties have been modified over a large area.

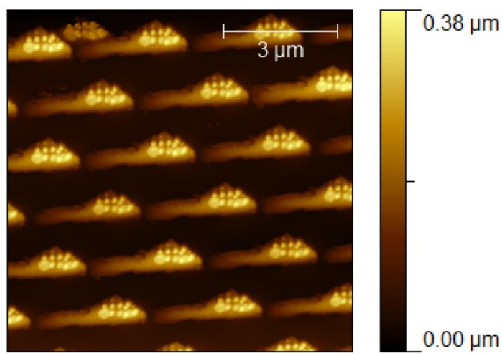

FIG. S3. AFM image of test structure TGT 1 shows the existence of two rows of spikes grown on the tip end. The distance between spikes is 200 nm.

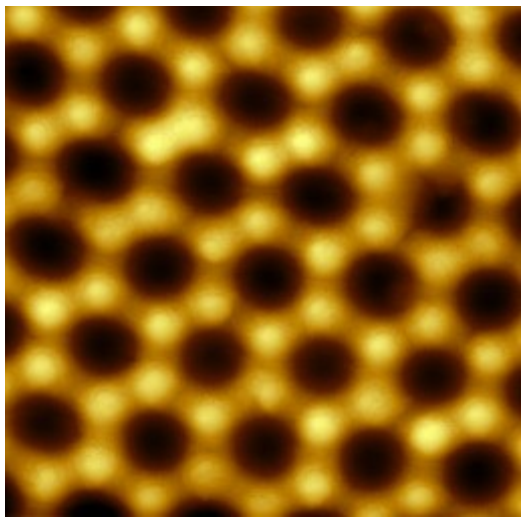

FIG. S4. Test of tip durability. The movie consists of 100 scans (500×500 nm$^2$) of a dimpled aluminum sample (multimedia view Al_101.wmv).